\documentclass{article}

\usepackage{arxiv}

\usepackage[utf8]{inputenc} 
\usepackage[T1]{fontenc}    
\usepackage{hyperref}       
\usepackage{url}            
\usepackage{booktabs}       
\usepackage{amsfonts}       
\usepackage{nicefrac}       
\usepackage{microtype}      
\usepackage{lipsum}		
\usepackage{graphicx}
\usepackage[numbers]{natbib}
\usepackage{doi}
\usepackage{multicol}
\usepackage{multirow}
\usepackage{subcaption}
\newcommand{\degree}{$^{\circ}$}
\newcommand{\private}{{\large\textbullet}}
\newcommand{\raw}{{\large\textopenbullet}}

\title{Exploiting the Uncoordinated Privacy Protections of Eye Tracking and VR Motion Data for Unauthorized User Identification}

\author{ 
    \href{https://orcid.org/0000-0002-7656-2662}{\includegraphics[scale=0.06]{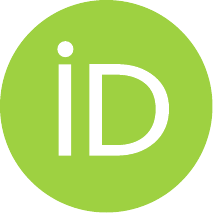}\hspace{1mm}Samantha Aziz} \\
	Department of Computer Science\\
	Texas State University\\
	San Marcos, Texas, USA\\
	\texttt{sda69@txstate.edu} \\
    \And	
    \href{https://orcid.org/0000-0001-7890-8842}{\includegraphics[scale=0.06]{orcid.pdf}
    \hspace{1mm}Oleg Komogortsev} \\
	Department of Computer Science\\
	Texas State University\\
	San Marcos, Texas, USA \\
	\texttt{ok@txstate.edu} \\
}

\date{}

\renewcommand{\shorttitle}{Exploring the Uncoordinated Privacy Protections of Eye Tracking and VR Motion Data for Unauthorized User Identification}

\hypersetup{
pdftitle={Exploring the Uncoordinated Privacy Protections of Eye Tracking and VR Motion Data for Unauthorized User Identification},
pdfsubject={cs.HC},
pdfauthor={Samantha Aziz, Oleg Komogortsev},
pdfkeywords={eye tracking, privacy, biometrics, user profiling},
}

\begin{document}
\maketitle

\begin{abstract}
Virtual reality (VR) sensors capture large amounts of user data, including body motion and eye tracking, that contain personally identifying information.
While privacy-enhancing techniques can obfuscate this data, incomplete privacy protections risk privacy leakage, which may allow adversaries to leverage unprotected data to identify users without consent.
This work examines the extent to which unprotected body motion data can undermine privacy protections for eye tracking data, and vice versa, to enable user identification in VR.
These findings highlight a privacy consideration at the intersection of eye tracking and VR, and emphasize the need for privacy protections that address these technologies comprehensively.
\end{abstract}

\keywords{privacy \and eye tracking \and virtual reality \and biometrics \and user profiling \and user identification}

\section{Introduction}

Virtual reality (VR) systems collect a large amount of data from a variety of sensors, including those that track the position and movement of a user's body.
Sensors such as Inertial Measurement Units (IMUs) track a user's motion over time, which enables many of VR's core functionalities.
However, the collection and utilization of such data may inadvertently reveal more information about a user than what is necessary for a given application.
For example, while motion tracking data is useful for delivering an immersive user experience, it also can also convey physical traits and behavioral patterns that may be used to infer additional information about the user.
The potential misuse of motion tracking data to reveal an excessive amount of information about a user raises concerns for user privacy, as a user may be profiled or even re-identified across play sessions without their knowledge. 

In addition to sensors that track the user's body in a virtual space, eye tracking technology is becoming an increasingly common feature on VR platforms.
Eye tracking enables several functions in VR, including the delivery of power savings through foveated rendering~\cite{Patney}, facilitating the correction of visual distortions that contribute to VR sickness~\cite{Martschinke2019}, and improving the perceived realism of virtual avatars during social interactions~\cite{garau2003impact}.
In addition to delivering these functionalities to VR platforms, eye tracking introduces additional concerns for user privacy in VR platforms.
Like motion data in VR, eye tracking data also captures sensitive information about a user that can be used to infer personal attributes~\cite{kroger}.
However, whereas motion data in VR may convey information about a user's physical traits and kinesthetic behavior, eye tracking can more directly reveal a user's cognitive behavioral traits such as their personality or interests.
If eye tracking data is misused to reveal sensitive characteristics about a user, these insights can complement the rich body of user information that can already be inferred from VR motion data. 

While much research has been dedicated to investigating the privacy implications of eye tracking and VR motion data separately, relatively fewer works address the privacy implications that may arise when these data streams are available simultaneously.
The combination of these data types could increase the potential for privacy violations in ways that have not yet been fully understood.

To address this gap, we explore how the simultaneous availability of these data streams can introduce privacy vulnerabilities, especially when existing privacy-enhancing techniques fail to account for the combined risk of user identification across multiple sensors.
Using a publicly available data set of eye tracking, VR headset, and VR controller motion data, we demonstrate that VR motion data and eye tracking signals can be used to identify users in VR platforms, and examine the extent to which the application of privacy-enhancing mechanisms to this data can reduce the risk of user identification.
We then simulate an adversarial privacy attack where data from VR motion sensors is used to circumvent the privacy protections applied to eye tracking data, and vice versa.
Our results show that without comprehensive privacy safeguards addressing all available data sources, these privacy mechanisms can be bypassed to enable user identification through unprotected data streams.
These findings underscore the need for comprehensive privacy solutions that can account for the novel privacy risks of multi-sensor environments in VR.
This study contributes to a deeper understanding of how the privacy considerations for VR motion data and eye tracking data interact, and provides motivation for future work that develops comprehensive privacy solutions that are suitable for use in multi-sensor VR environments.

\section{Background and Related Work}
\label{sec:prior}
When using a VR platform, user activity is captured and recorded by a variety of sensors.
VR platforms capture user activity using a variety of sensors, including IMUs, proximity sensors, and cameras. 
In addition to fulfilling benign functionalities, this activity data can also convey information about the user that is not strictly necessary for a VR application to have knowledge of; for example, the average height of the headset in the user's world space over time can reveal the user's approximate height~\cite{Nair2023_incognito}.
The information that can be inferred about a user is not limited to physical characteristics like body proportions; VR data can also be analyzed to secondarily reveal personal characteristics such as gender~\cite{Tricomi, Gil2023,Nair_beat} age~\cite{Tricomi, Gil2023, Nair_beat}, and socioeconomic traits that are correlated with these characteristics~\cite{Nair_beat}.
Taken together, this information can be used to build a virtual profile of a user based on their motion data.

Several works in literature demonstrate that users can be uniquely identified through VR motion data---this data conveys information about the user's physical characteristics~\cite{Nair_beat} and behavioral patterns~\cite{Wierzbowski2022} that can be used to re-identify them across play sessions in a virtual environment~\cite{Miller2020}. 
Users can also be re-identified across several play sessions or uniquely identified among a population of other users~\cite{Meng2024}.
The ability to covertly extract personally identifying information from VR motion data has significant implications for user privacy, even if a user's real-world identity is not always attached to their virtual persona.
Although a privacy vulnerability of this kind is concerning in any application, it becomes particularly problematic in emerging VR use cases where data is gathered from sensitive populations, such as children~\cite{cao2024}, or applied in privacy-sensitive contexts such as delivering healthcare~\cite{lake} or education~\cite{kaur}.

Efforts to protect user privacy in VR often manifest as system-level security protections that prevent unauthorized access to user-generated data, such as encryption, firewalls, and intrusion detection~\cite{kulal2022security,de2019security}.
Although these techniques can help prevent data from being accessed through unambiguously malicious activities such as exploiting security vulnerabilities, these techniques do not prevent privacy violations that can be committed using data that is willingly provided by users.
When VR motion data is provided to a malicious application, it is possible for these privacy violations to take place before the privacy risk is recognized by either the system or the user.
For example, a VR application that is granted motion data for a benign purpose may secondarily use it to infer sensitive information about the user.
Nair et al.~\cite{nair2023exploring} explore one such scenario by designing a VR game that obstinately collects motion data to enable gameplay, but which covertly elicits motion data that reveals sensitive user information.

Policy-based solutions that dictate how VR data should be used may also be insufficient for protecting users, because VR applications may not adhere to the data management practices stated in their privacy policies.
After analyzing the network traffic of 140 popular VR applications, Trimananda et al.~\cite{Trimananda2022} observe that 70 percent of data transmissions were inconsistent with their respective application's privacy policy, and often transmitted personally sensitive information that could be used for user identification. 
These gaps in security-based and policy-based solutions suggest that additional mechanisms are needed to protect users from being profiled from their VR motion data.

To fulfill the need for privacy enhancement in VR, several works have proposed privacy-enhancing techniques that make personally identifying information more difficult to extract from user-generated data.
Rather than litigating access to (or utilization of) potentially sensitive data, these techniques apply data processing techniques to user-generated data streams to obfuscate personally identifying details before they are made accessible to third parties. 
Preliminary efforts by Sun et al.~\cite{Sun2024} suggest that differential privacy can be applied to obscure behavioral patterns that may identify a user; however, this investigation was only performed on simulated human activity.
Wierzbowski et al.~\cite{Wierzbowski2022} observe that eliminating differences in users' characteristics (e.g., by standardizing all users' heights by seating them in a chair) while collecting VR motion data can reduce user identification rates; this places constraints on how users can interact with their devices, which makes this method difficult to apply in a practical setting. 
Nair et al.~\cite{Nair2023_incognito} apply differential privacy to obfuscate physical characteristics from existing data without restricting how users interact in VR. 
Similarly, Meng et al.~\cite{Meng2024} propose adding noise to user motion data before it is used to render the virtual avatar's motion, thereby obscuring the user's movement patterns in a virtual space. 

The privacy landscape for VR is further complicated when considering the growing integration of eye tracking technology in VR platforms. 
Eye tracking is used to provide vital functionalities to VR platforms, such as power savings via foveated rendering~\cite{Patney} and correction of visual distortions that can cause discomfort to users~\cite{Martschinke2019}.
However, like other sensors that capture human activity, eye tracking also introduces concerns for user privacy by enabling user profiling and identification.
Eye tracking data is a particularly rich source of information that gives insights into a user's cognition.
It can serve as a behavioral biometric that uniquely identifies users~\cite{Lohr2022}, and can be used to infer information related to the user's cognitive state, such as their fatigue level~\cite{fatigue}, cognitive load~\cite{cognitive}, and emotional state~\cite{emotion} and personality type~\cite{personality}.
Several personal characteristics that can be inferred through VR, such as gender and age, can also be inferred through eye tracking data~\cite{kroger}. 

As eye tracking becomes an increasingly standard feature in consumer-facing VR devices---including Meta's Quest Pro, HTC's Vive Pro Eye, and Apple's Vision Pro---privacy concerns regarding the use of this data have prompted the development of privacy-enhancing techniques for gaze data on these devices.
Efforts to enhance privacy in eye tracking include system-level techniques to restrict the transmission of sensitive data~\cite{Hu2022_otus, Bozkir2020_estimation, davidjohn2021}, and the application of data-augmenting privacy mechanisms to remove sensitive information from gaze data while allowing it to be used for benign purposes, for both real time gaze interaction~\cite{Wilson2024,Li2021} and during post-hoc data analysis of previously collected gaze data~\cite{davidjohn2021, Steil2019b, Liu2019}.

The privacy-enhancing techniques discussed thus far were developed for VR motion data or eye tracking data in isolation, under the assumption that no other potentially sensitive data streams are available to the adversary. 
Although the privacy implications of VR and eye tracking remain open areas of research on their own, it is becoming increasingly important to address the privacy implications that arise when they are used together.
The joint integration of motion sensors and eye tracking on consumer-facing VR platforms is making the availability of both sources of data collected synchronously increasingly common, which can be problematic for user privacy if not acknowledged.
These data streams can potentially provide complementary information that increase the accuracy of user identification; for example, an adversary may build a more complete profile of a user by combining complementary information about their physical bodies (extracted from VR motion) with information about the way they think, act, and allocate their attention (extracted from eye tracking data). 
Several works show that analyzing gaze behavior in conjunction with VR motion data can contribute to more successful user identification~\cite{Liebers2021, Pfeuffer2019, Wierzbowski2022, Olade2020}, but this trend has not been discussed in the context of its implications for user privacy.

Given that privacy protection strategies for eye tracking and VR data have been developed in isolation, additional privacy considerations may emerge when these technologies intersect.
In this work, we explore one such consideration where an adversary attempts to identify users on a VR platform, circumventing incomplete privacy protections by leveraging data streams collected by sensors for VR motion tracking and eye tracking.
While the potential privacy considerations of combining sensor data in VR platforms has been discussed conceptually in prior research~\cite{Munilla_Garrido_2024, Giaretta2024}, this investigation empirically explores the impact of applying privacy mechanisms on user identification rates when VR motion and eye tracking data are available simultaneously.
First, we empirically evaluate the effectiveness of two state-of-the-art privacy mechanisms designed for eye tracking and VR motion data in mitigating privacy leakage from this data, measured by the reduction of user identification rates.
We also explore how incomplete privacy protections affect user identification rates by simulating a plausible scenario where privacy protections are applied only to a subset of the available data streams; in doing so, we demonstrate that incomplete privacy protection can enable unauthorized user identification.

\section{Threat Model}
\begin{figure*}
    \centering
    \includegraphics[width=\linewidth]{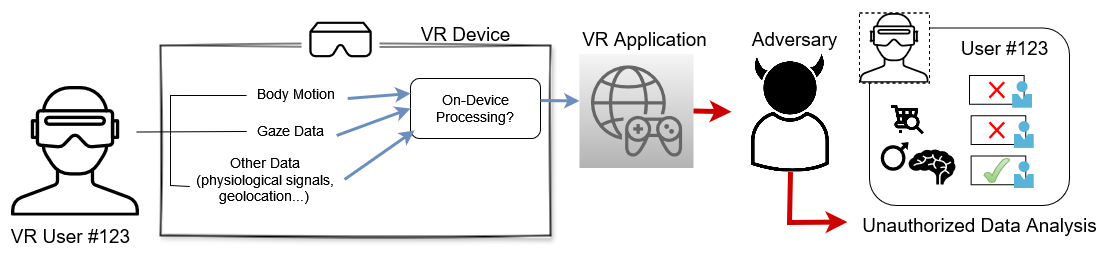}
    \caption{An illustration of the threat model proposed for this investigation, where a malicious third-party developer attempts to identify users using a combination of data streams collected from a user's VR device. Blue arrows represent a benign flow of information through this system, and red arrows show malicious information flow.}
    \label{fig:threat}
\end{figure*}

For this investigation, we explore the problem of unauthorized user re-identification in VR platforms.
The threat model for this work, illustrated in \ref{fig:threat}, involves an adversary who attempts to perform user identification using body motion and gaze data collected by a VR device. 
This adversary does not necessarily engage in malicious activity to gain access to these data streams; for example, if they are a developer for an application in VR, and could be granted access to body telemetry and gaze data streams to enable the functionality of their otherwise benign application~\cite{nair2023exploring}. 
As such, security measures that detect unusual data transmission or unauthorized access to information would not necessarily prevent this attack from occurring. 

Given access to this data, the adversary then attempts to infer user identities using the signals that they collected from the device.
The adversary may or may not have knowledge of which data streams (if any) had privacy mechanisms applied to them before they were provided by the device.
Through repeated experimentation, they may even be able to identify which data streams are unprotected, and use this information to inform future attacks. 
We explore whether this adversary would be able to successfully perform user identification when (a) none of the data streams have had privacy protections applied, (b) all of the data streams have had privacy protections applied, and (c) when some, but not all, data streams had privacy protections applied. 

\section{Methodology}
Given that gaze data and body motion data is often collected synchronously for benign purposes in VR, we first seek to evaluate whether combining these data streams increases the accuracy of user identification.
We then investigate whether these data streams could be used to circumvent privacy-preserving techniques meant to prevent user identification, when only a subset of data undergoes privacy enhancement.
We performed an empirical study where we attempted user identification on a publicly available data set of eye movement and body motion data collected in VR.

\subsection{Data Set}
We utilized OpenNEEDS~\cite{Emery2021}, a data set of gaze, head, and hand signals that were collected from 44 participants while they freely explored indoor and outdoor scenes in VR. 
This data was collected in two sessions, where participants were instructed to explore their virtual environment and interact with various objects at their discretion.
During each session, gaze data was captured using a custom eye tracker embedded in a VR headset, and motion tracking data was captured for the head, left hand, and right hand using equipment from an Oculus Rift. 
All data was captured at a sampling rate of 90 Hz.
We excluded participants from this investigation if they did not have at least two sessions' worth of data, or if they had fewer than 15 seconds' worth of data in either session.
As a result, we utilized data from 38 participants total.

\subsection{Privacy Mechanisms}
We selected two privacy mechanisms that have been previously proposed in literature for this investigation; these mechanisms have demonstrated the ability to enhance user privacy for gaze and VR head motion data, respectively.
However, the two have not been tested in conjunction with one another. 
These mechanisms were selected because they are feasible to implement in real-time and produced favorable privacy-utility trade-offs in their original publications, which are important requirements to ensure that privacy protections do not interfere with the system's overall usability.

We applied each privacy mechanism using the strongest privacy settings recommended in each paper. 
While privacy settings are typically fine-tuned to suit an individual data set, we found that the privacy budgets proposed by each method were appropriate to use with our data.
In this section, we give a brief overview of each of these mechanisms.

\subsubsection{Gaze Data Privacy Mechanism}
We employ privacy mechanism proposed by Wilson et al.~\cite{Wilson2024} for gaze data, which obfuscates personally identifying information in gaze data by applying a smoothing operation to the gaze data as it is streamed from the device.
To smooth the data, each gaze sample is re-defined as the linearly weighted average of $B$ preceding samples.
For this investigation, we set the parameter $B = 108$, which corresponds to the 1.2 seconds of data that the original work was validated on.

\subsubsection{Head and Hand Data Privacy Mechanism}
We employ a portion of MetaGuard from Nair et al.~\cite{Nair2023_incognito} to serve as our privacy mechanism for our headset and hand data. 
MetaGuard applies $\epsilon$-differential privacy to obscure physical characteristics of the user's body, thereby reducing the chance that users are re-identified based on their physical traits. 
We utilize the portion of their method that obfuscates the users' height and wingspan by adding bounded Laplacian noise to the vertical component of headset data and to the positional hand data, respectively. 
We estimate the upper and lower bounds of the noise distributions by approximating the height of the users in OpenNEEDS.
To do so, we designate the height and wingspan of each user as the average vertical component of their headset data during their first recording, assuming a 1:1 height-to-wingspan ratio.
Using this method, we then constrain the range of both heights and wingspans in our data to [1.32, 1.82] meters and apply bounded Laplacian noise with these bounds.
Following the privacy budgets proposed in the original work, we set $\epsilon = 1$ for headset data and $\epsilon = 0.5$ for hand data.

We do not perturb the horizontal or depth components of our headset data, because the original method only does so to prevent an adversary from inferring the dimensions of the user's play area---because all of our users utilized the same space and the justification for perturbing these components has little relevance for personal identity, we determined that perturbing these data streams would not be necessary for our problem setting.

For this investigation, we enforce parity on the privatization statues of VR motion data---in other words, if head and hand data are both available, they are either both privatized, or both unprivatized.

\subsection{Biometric Authentication Model}
\label{sec:ekyt}
We employ EyeKnowYouToo (EKYT)~\cite{Lohr2022}, the state-of-the-art biometric model for eye movement biometrics, as our biometric authentication model.
While EKYT was originally developed and validated for gaze-based biometrics, we extend EKYT to perform biometric authentication using headset and hand motion data by adding them as additional channels to the input. 

To test the effect of combining data streams on biometric performance, we train and evaluate 7 models on unprivatized (unmodified) data. These models establish a baseline of biometric performance for this investigation, and their results are evaluated in Section~\ref{sec:no-privacy} (Experiment IDs 01-07). 
We train and evaluate an additional 7 models on fully privatized data to establish identification rates under the best-case privacy protection that we can expect from this investigation.
In this context, this means that every available data stream has been processed by its respective privacy mechanism.
These models are evaluated in Section~\ref{sec:all-privacy} (Experiment IDs 08-14).

To test the ability of the biometric model to circumvent incomplete privacy protections with other sources of data, we train and test an additional 6 models on various combinations of privatized and unprivatized data steams, where at least one data stream is not privatized.
These models investigate the extent to which EKYT can leverage alternative sources of data to circumvent privacy protections, and are evaluated in Section~\ref{sec:some-privacy} (Experiment IDs 15-20). 

\subsubsection{Data Pre-Processing}
\label{sec:preprocess}
Eye movement data was processed according the methodology described in EKYT's original implementation~\cite{Lohr2022}.
Given the positional gaze data provided by the VR device, we first calculate the velocity of the eye movement using a Savitzky-Golay differentiation filter with an order of 2 and a window size of 7. 
The velocity data is then split into non-overlapping windows of 5 seconds, which corresponds to 450 samples at 90 Hz. 
In each window, these velocities were clamped to realistic movement speeds of $\pm$1000\degree/sec, then normalized using the mean and standard deviation of eye movement velocities derived from the training data and replaced NaN values with 0. 
These processed velocity windows were then provided to EKYT for training and evaluation.
As a result, EKYT is provided windows of 450 samples, where each sample contains the horizontal and vertical velocity of the eye at a point in time, denoted as $G_t = (v_x,v_y)_t$.

During experimentation, we observed no particular benefit of extensively pre-processing VR motion data.
VR motion data was split into non-overlapping windows of 5 seconds and provided to the model for training and evaluation.
The model received 450-sample windows of positional headset data, where each sample contained the horizontal, vertical, and depth position of the headset in world space, denoted as $H_t = (h_x,h_y,h_z)_t$.
Although we have rotational data available from the headset as well, we chose to only provide headset positions as input in order to prevent information from any single data stream from being overrepresented in the inputs. 
Hand data was also provided in 450-sample windows, with each sample containing $(L, R)_t = (\{l_x,l_y,l_z\},\{r_x,r_y,r_z\})_t$, which respectively denote the horizontal, vertical, and depth of the left and right hands in world space at time $t$.
We only provide positional data of the hands to the model as input, but we acknowledge that including rotational data and interaction patterns (e.g., button presses, gestures) may also contain information that can uniquely identify a user~\cite{Pfeuffer2019}.

Each data stream is provided to EKYT as an independent, but time-correlated channel to the input; because we were interested in observing the impact that each data stream had on overall biometric performance, we did not explore sensor fusion techniques that might have made it easier for EKYT to extract complementary information across these data streams. 

\subsubsection{Training and Evaluation}

We train and test separate authentication models for each test condition described in Section~\ref{sec:ekyt}.
Each model was trained for 100 epochs, with a learning rate of 3e-4.
Due to the limited number of subjects in our dataset, each model is trained using four-fold cross-validation.
Each fold is trained on 75\% of subjects and evaluated on the remaining 25\% of subjects that were held out, ensuring that the training and evaluation sets are subject-disjoint.
We report each model's biometric performance as the mean authentication performance on the held-out subjects across all four folds.

We evaluate biometric authentication performance in two settings---biometric verification and biometric identification. 
During biometric verification, EKYT receives both a sample of data and a claimed user identity, and compares that sample to the user's enrollment template to determine whether the sample could have been generated by the claimed identity. 
The system then decides whether the sample is sufficiently similar to the enrollment template, returning ``True'' if the samples match and ``False'' otherwise. 
We summarize biometric verification performance using Equal Error Rate (EER). 
Generally, lower EER indicates better biometric performance.
During biometric identification, EKYT receives a sample of data and returns the identity of the user who most likely generated the data. 
We report biometric identification performance as the Rank-1 Identification Rate (Rank-1 IR), or the percentage of times that EKYT returns the correct identity. 
Generally, higher Rank-1 IR indicates better biometric performance.

In an ideal scenario, data that has undergone privatization would not contain any identifying information.
A biometric model would therefore produce chance-level biometric performance when using this data, which corresponds to 50\% EER for verification and $1/38 \approx 2.6\%$ IR for identification.
To assess the effect of privatization on our data set, we first establish baseline biometric performance when using combinations of unmodified gaze, head, and hand data, then then observe how that biometric performance changes after applying privacy mechanisms to different data streams.
We then assess how the privacy gains afforded by these mechanisms may be circumvented by only applying privacy mechanisms to a subset of available data, and allowing EKYT access to all other unmodified data streams.

\section{Results}
\subsection{Combining Unmodified Data Streams for User Authentication}
\label{sec:no-privacy}
We first establish a baseline for EKYT's biometric authentication performance when it is trained and tested on unmodified data from OpenNEEDS.
The authentication rates observed from unmodified data, summarized in Table~\ref{tab:unmodified} and Figures~\ref{fig:bar-private} and~\ref{fig:roc-effect-of-privatization}, represents the risk of user identification in a scenario where a VR platform streams sensor data to third party applications without applying privacy protections.
We present biometric performance when EKYT has access to a single data stream (E01-E03), two data streams (E04-E06), and all three data streams (E07) to explore the effect of combining unmodified data streams on user authentication rates.

We first observe that providing a single data stream (E01-E03) produces authentication rates that are well above chance-level performance, which indicates the EKYT can meaningfully extract user-specific information that can be used to re-identify them across play sessions. 
Gaze data (E01) and headset position data (E02) produce the similar rates of user authentication, and yield similar verification and identification rates on average.
On the other hand, it appears that hand positions (E03) were not as informative for inferring user identity; we observe a notable decrease in biometric performance across both authentication settings relative to gaze and headset data. 
While features that are related to hand motion data have shown promise for use as a soft biometric (e.g., hand dominance and derivable body proportions~\cite{Pfeuffer2019}), there is relatively less evidence that raw hand motion data alone is as suitable for user identification as gaze data~\cite{Lohr2022, 9555831} and headset data~\cite{liebers2021-behavioral-normalization, Wierzbowski2022}.

We next observe the effect of combining two sources of data on user authentication rates.
Overall, combining two data sources does not appear to improve biometric performance over utilizing any single data stream, but it can help overcome poor biometric performance when a more informative data stream (e.g., gaze or head) is supplied to the model alongside a less informative data stream (e.g., hands).
For example, combining head and hand data (E04, known hereafter as ``VR motion data'') on authentication rates yields biometric authentication performance similar comparable to the test condition where only headset data is available (E02), which is a notable improvement in performance relative to the hands-only condition (E03).
Similarly, combining gaze and hand data produces a negligible degradation in biometric performance on average relative to the gaze-only condition (E06 versus E01), indicating that EKYT can effectively leverage the more informative data stream to meaningfully distinguish between users. 
We also observe a slight (but likely not significant) deterioration in authentication performance between E01/E02 and E05 after combining gaze and head data, indicating that EKYT did not distinguish users more meaningfully on average when these data sources were combined.

Finally, combining all three data streams (E07) does not appear to significantly affect verification rates over the best individual data stream, but does improve the identification rate to its highest observed value of 75\%. 
The modest improvement in identification rate suggest that combining data streams may give the biometric model more opportunity to identify unique behavior across all three streams compared to using a single data stream in isolation, as person-specific idiosyncrasies would be captured simultaneously in all three sources of data.

\begin{table}[tb]
  \caption{Biometric performance of EKYT when trained and tested on unmodified data streams. Performance is averaged across 4-fold validation.}
  \label{tab:unmodified}
  \scriptsize%
	\centering%
  \begin{tabular}{%
    l
	*{3}{c}%
    *{2}{r}%
	}
  \toprule
   ID & Gaze &   Head &   Hand &   EER (\%, $\downarrow$) &   Rank-1 IR (\%, $\uparrow$) \\ 
  \midrule
  E01 & \raw &  - & - & 18.9 $\pm$ 8.5 & 62.5 $\pm$ 17.1 \\ 
  E02 & - & \raw   & - & 18.6 $\pm$ 2.1 & 60.0 $\pm$ 21.6 \\ 
  E03 & - & -  & \raw  & 30.8 $\pm$ 7.4 & 42.5 $\pm$ 22.2 \\ 
  E04 &  - & \raw  & \raw  & 19.2 $\pm$ 6.4 & 65.0 $\pm$ 20.8 \\ 
  E05  & \raw & \raw   & - & 22.5 $\pm$ 12.6 & 62.5 $\pm$ 28.7 \\ 
  E06  & \raw & -  & \raw  & 25.0 $\pm$ 5.8 & 57.5 $\pm$ 20.6 \\ 
  E07 & \raw & \large\raw   & \raw  & 19.4 $\pm$ 7.7 & 75.0 $\pm$ 12.9 \\ 
  \bottomrule
  \multicolumn{6}{l}{(\raw)~Unmodified data, (\private)~Privatized data, (-)~Data not used}
  \end{tabular}%
\end{table}

\subsection{User Authentication with Fully Privatized Data}
\label{sec:all-privacy}
\begin{figure}
    \centering
    \includegraphics[width=0.6\linewidth]{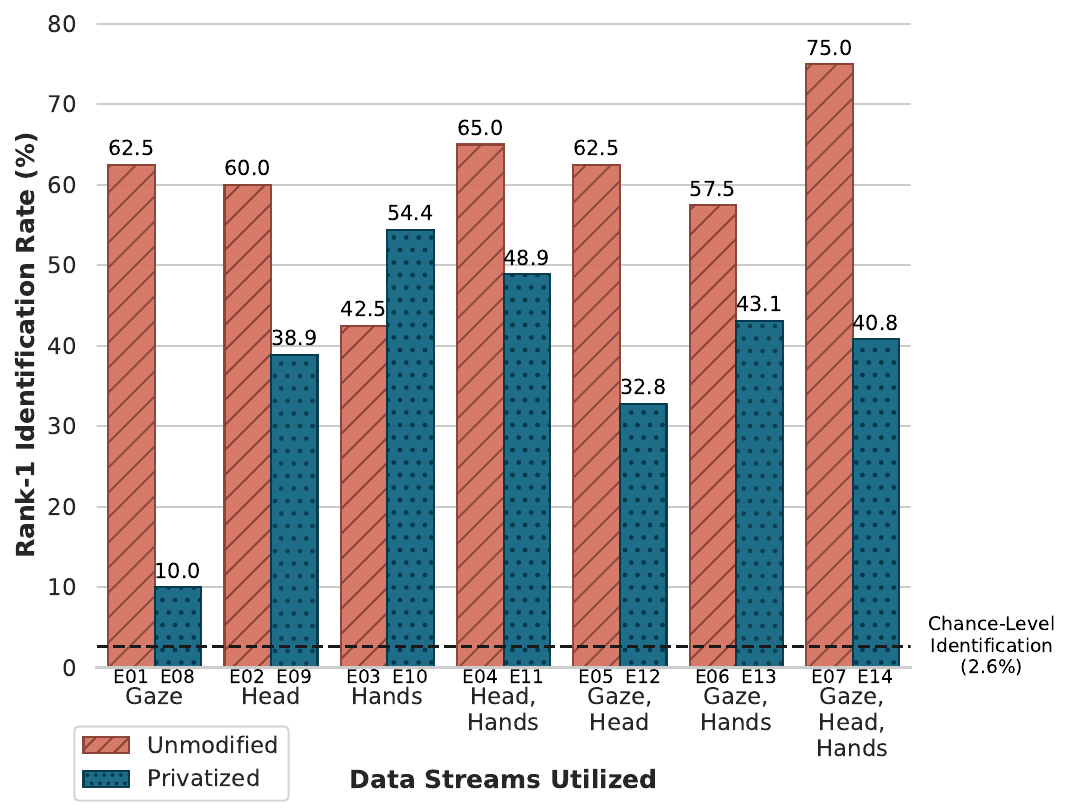}
    \caption{A visualization summarizing the effect of privatization on biometric identification rates.}
    \label{fig:bar-private}
\end{figure}

\begin{figure}
    \centering
     \begin{subfigure}[b]{0.235\textwidth}
         \includegraphics[width=\textwidth]{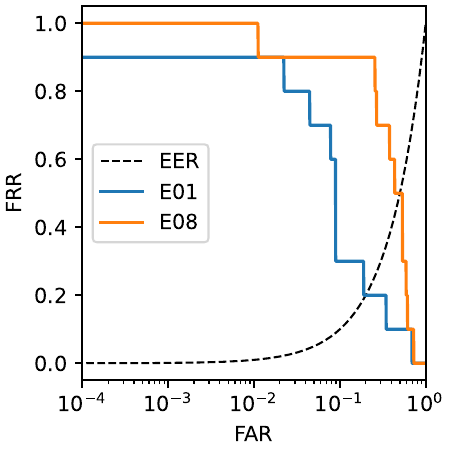}
         \caption{Gaze data.}
         \label{fig:roc-gaze}
     \end{subfigure}
      \begin{subfigure}[b]{0.235\textwidth}
        \includegraphics[width=\textwidth]{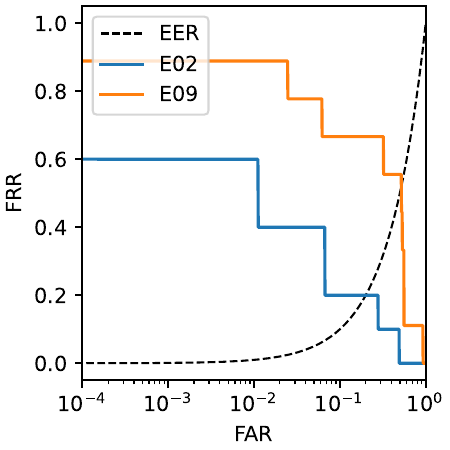}
         \caption{Headset data.}
         \label{fig:roc-head}
     \end{subfigure}
      \begin{subfigure}[b]{0.235\textwidth}
        \includegraphics[width=\textwidth]{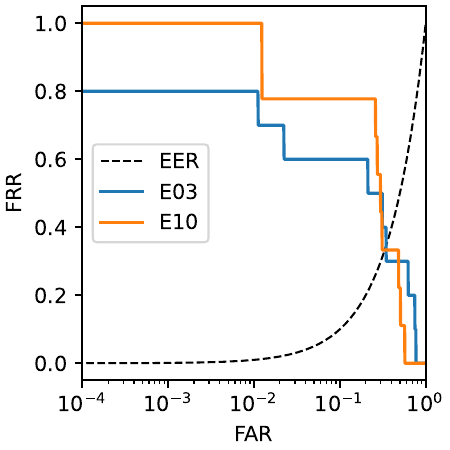}
         \caption{Hand data.}
         \label{fig:roc-hands}
     \end{subfigure}
      \begin{subfigure}[b]{0.235\textwidth}
        \includegraphics[width=\textwidth]{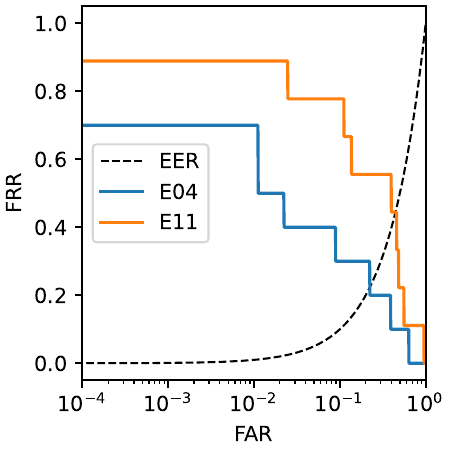}
         \caption{VR motion data.}
         \label{fig:roc-vr}
     \end{subfigure}
    \caption{ROC curves comparing the effect of privatization on biometric verification rates for each data stream individually, and for VR motion data.}
    \label{fig:roc-effect-of-privatization}
\end{figure}

We next explore a best-case interaction scenario where all of the data streams that we consider have undergone privacy enhancement.
Specifically, we explore the impact of applying Wilson et al.~\cite{Wilson2024}'s smoothing mechanism to the gaze data and applying Nair et al.~\cite{Nair2023_incognito}'s MetaGuard mechanism to VR motion data, respectively.
When applied to their respective data streams, these privacy mechanisms obfuscate the personal information encoded in the data, producing ``privatized'' data is ideally absent of personally identifying features.
This hinders EKYT's ability to learn meaningful user-specific patterns during training, thereby reducing the chances that users can be re-identified across play sessions.
In an ideal scenario where the privatized data is totally absent of personally identifying information, the resulting biometric authentication rates would be at or near chance-level performance, which corresponds to 50\% EER and 2.6\% IR for verification and identification, respectively.  
We observe the impact of privatization on degrading EKYT's biometric authentication rates toward this ideal.
Table~\ref{tab:private} shows the average biometric performance of EKYT models that were trained on privatized data, while Figures~\ref{fig:bar-private} and~\ref{fig:roc-effect-of-privatization} show biometric performance for identification and verification, respectively.

First, we evaluate the effectiveness of each privacy mechanism at reducing authentication rates for each data stream individually. 
The gaze-based privacy mechanism (E08) is the most effective at obscuring personal information from EKYT, degrading biometric performance closest to chance-level verification rates (47.2\% EER versus 50\% at chance) and identification rates (10\% IR versus 2.6\% at chance), as well as producing the greatest reduction in authentication rates relative to its unmodified counterpart (E01).
The privacy mechanism for head data produces relatively more modest privacy enhancement, degrading authentication performance from 18.6\% EER and 60\% IR (E02) to 31.7\% EER and 38.9\% IR (E09).
This degradation in authentication performance indicates that EKYT is less capable of recognizing users across sessions, which represents an increase in privacy relative to the unmodified data. 
On the other hand, the privacy mechanism applied to the hand data is less effective at reducing authentication rates, and therefore produces comparable (or even slightly better) biometric performance relative to unmodified data (E03 versus E10).
We theorize that the limited privacy protections we observe for our hand data stems from unique characteristics of both EKYT and our privacy mechanism of choice; while MetaGuard is designed obscures the physical properties that can be derived from a user's hand data, such as their wingspan and arm length~\cite{Nair2023_incognito}, it does not obscure motion patterns that can be used for identification on a short timescale~\cite{Miller2022}.
Because EKYT learns to extract such behavioral patterns by leveraging minute changes in gaze velocity trajectory~\cite{Lohr2022}, it is possible that EKYT can overcome our implementation of MetaGuard by extracting user-specific information from behavioral patterns rather than from physicality. 
It it also possible that the privacy budget employed for MetaGuard is not an optimal choice for the OpenNEEDS data set, as it was derived for another data set that utilized a different set of stimuli~\cite{Nair2023_incognito}.

Overall, the biometric performance when combining privatized data streams (E11-E14) is notably worse than their unmodified counterparts (E04-E07), which indicates some privacy enhancement is afforded when all of the available data streams undergo privacy enhancement. 
However, authentication performance for these combined data sources tend to reproduce the performance of the least effectively privatized data stream.
For example, when provided with privatized gaze and hand data (E13), EKYT produces comparable biometric performance to E10, where only the poorly privatized hand data is available. 
While the increased privacy protection for gaze data prevents EKYT from learning user-specific information from the user's gaze, it did not degrade biometric performance when EKYT has access to other, less effectively privatized data streams, as user-specific information was still learnable from these alternative sources. 
From these results, it's reasonable to consider that comprehensive privacy defenses are only as effective as their ``weakest link,'' or least effectively privatized component.

\begin{table}[tb]
  \caption{Biometric performance of EKYT when trained and tested after applying privacy mechanisms to all data. Performance is averaged across 4-fold cross validation.}
  \label{tab:private}
  \scriptsize%
	\centering%
  \begin{tabular}{%
    l
	*{3}{c}%
    *{2}{r}%
	}
  \toprule
   ID & Gaze &   Head &   Hand &   EER (\%, $\downarrow$) &   Rank-1 IR (\%, $\uparrow$) \\ 
  \midrule
	E08 & \private &  - & - & 47.2 $\pm$ 4.7 & 10 $\pm$ 8.2 \\
    E09 &  - & \private   & - & 31.7 $\pm$ 9.3 & 38.9 $\pm$ 10.5 \\
    E10 &  - & -  & \private  & 26.7 $\pm$ 10.6 & 54.4 $\pm$ 23.5 \\
    E11 &  - & \private  & \private  & 29.1 $\pm$ 11 & 48.9 $\pm$ 20.8 \\
    E12 &  \private & \private   & - & 26.1 $\pm$ 5.8 & 32.8 $\pm$ 14.4 \\
    E13 &  \private & -  & \private  & 33.3 $\pm$ 8.2 & 43.1 $\pm$ 13.4 \\
    E14 &  \private & \private   & \private  & 28.8 $\pm$ 8.8 & 40.8 $\pm$ 22.4 \\

    \midrule
    \multicolumn{4}{c}{Chance-Level Performance} & 50 & 2.6 \\
  \bottomrule
    \multicolumn{6}{l}{(\raw)~Unmodified data, (\private)~Privatized data, (-)~Data not used}
  \end{tabular}%
\end{table}

\subsection{Circumventing Partial Privacy Protections for User Authentication}
\label{sec:some-privacy}
After establishing the upper and lower bound of user authentication rates for our data set (Sections~\ref{sec:no-privacy} and~\ref{sec:all-privacy}, respectively), we now investigate the privacy landscape that exists between these extremes.
In real-world VR applications, it cannot be assumed that comprehensive privacy protections are implemented or respected by every application, if these protections were not applied before the data was shared in the first place.
While the worst-case scenario is that there are no privacy protections in place, we are more likely to encounter inconsistently applied privacy protections across applications, where privacy protections have been applied to some potentially identifying data streams, but not to others.
The experiments in this section represent this patchwork approach to privacy enhancement, where some potentially sensitive data streams have undergone privacy enhancement prior to being shared and others are provided without any additional protections. 
We are interested in whether EKYT can leverage these unprivatized data streams to circumvent the privacy enhancements afforded by the privatized data, thereby defeating the purpose of privatization altogether. 
If this is the case, then the model would produce user authentication rates that are better the high-privacy scenario (E08-E14), and potentially reach comparable performance levels as the no-privacy scenario (E01-E07).

\begin{figure}
     \begin{subfigure}{0.33\textwidth}
         \includegraphics[width=.94\linewidth]{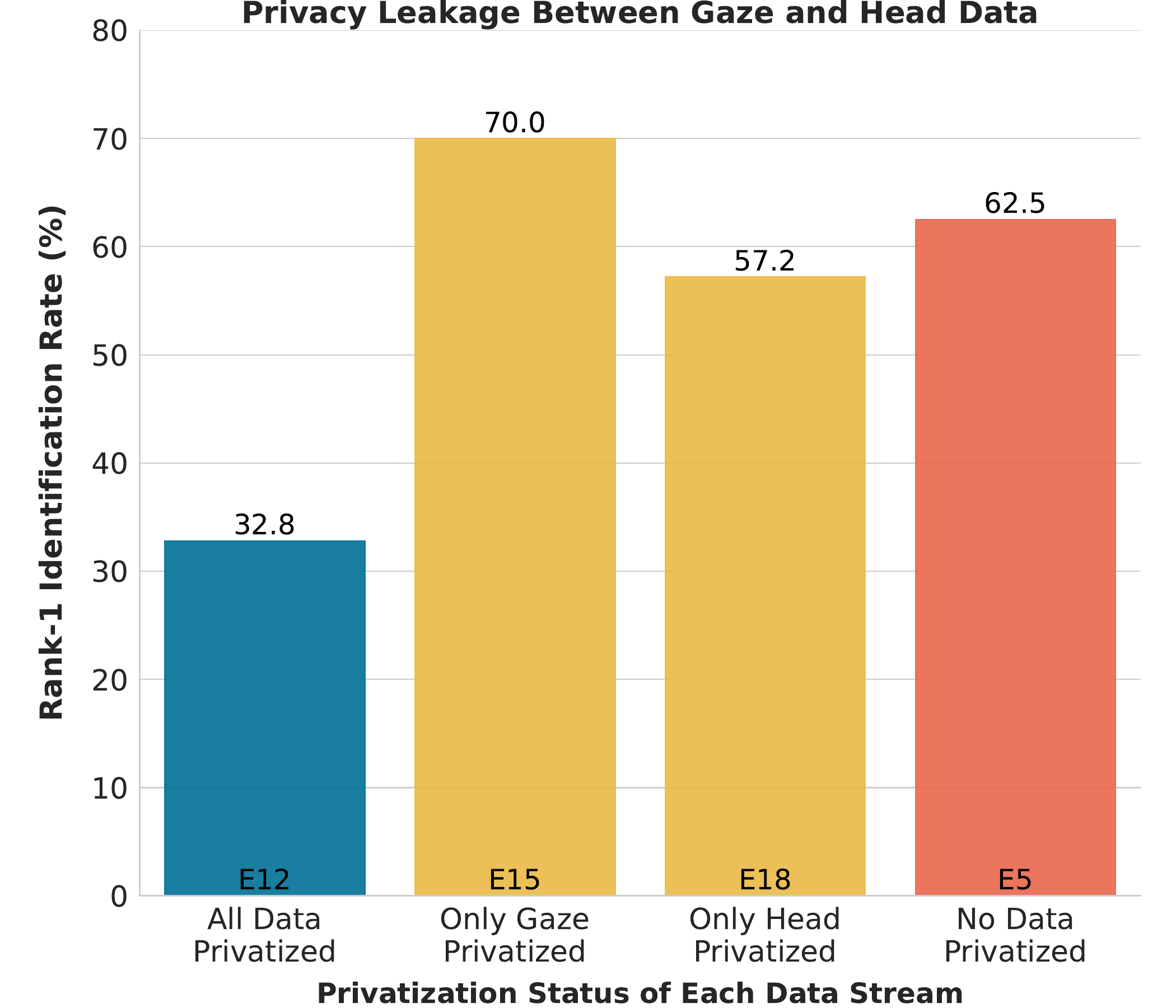}
         \caption{Gaze and head data.}
         \label{fig:leakage-gaze-head}
     \end{subfigure}
     \begin{subfigure}{0.33\textwidth}
        \includegraphics[width=.94\linewidth]{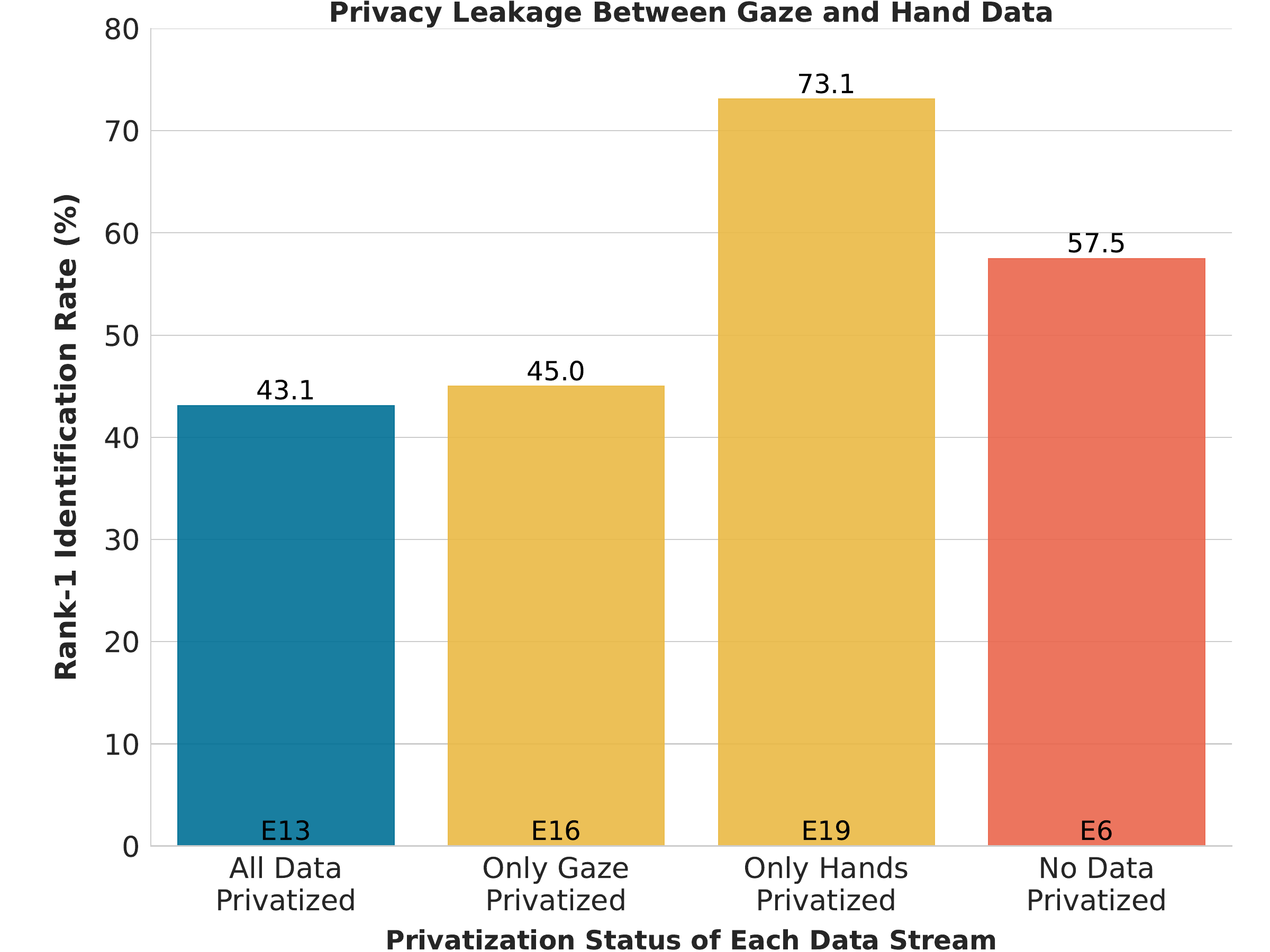}
         \caption{Gaze and hand data.}
         \label{fig:leakage-gaze-hands}
     \end{subfigure}
     \begin{subfigure}{0.33\textwidth}
        \includegraphics[width=.94\linewidth]{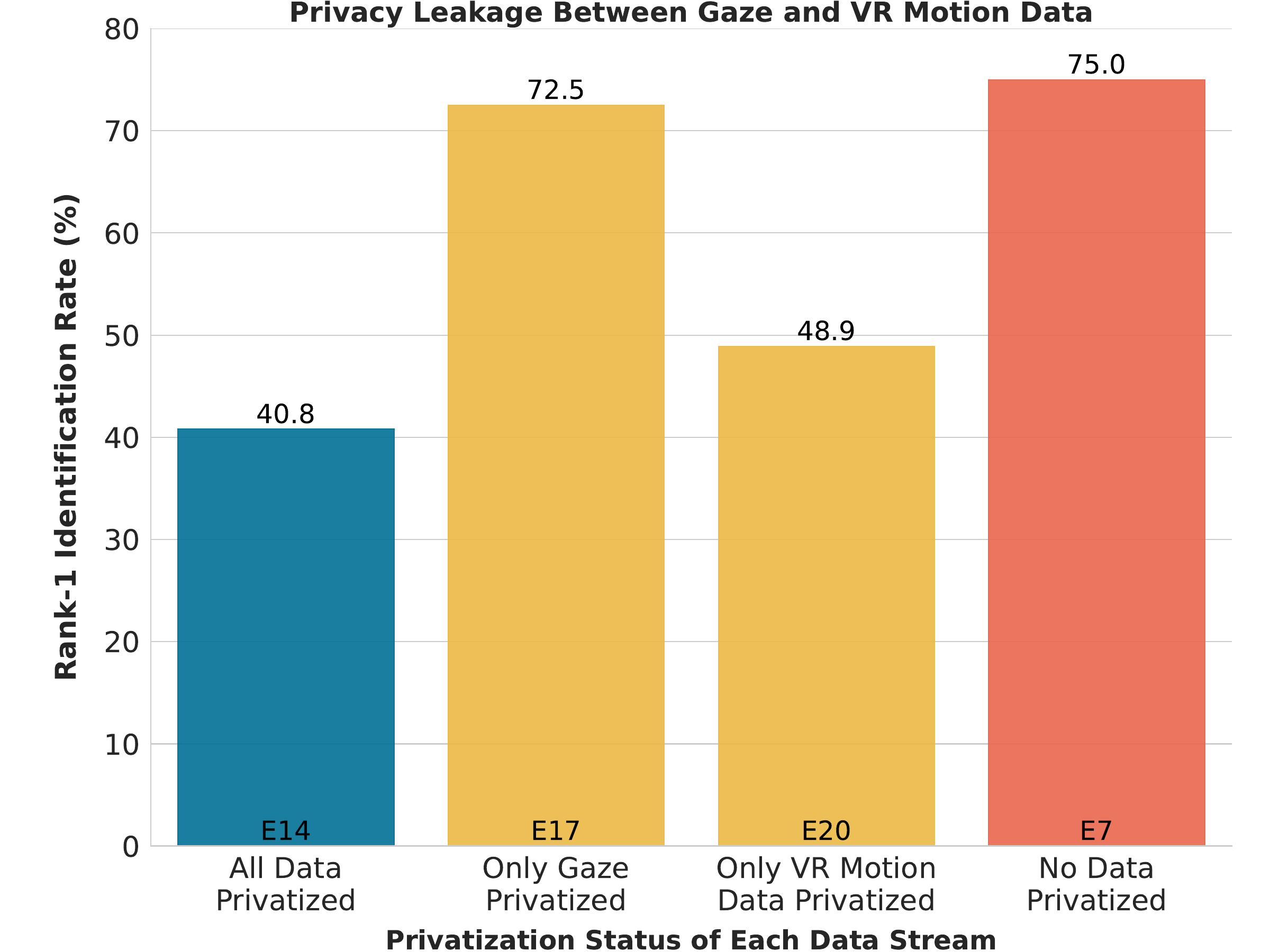}
         \caption{Gaze and VR motion data}
         \label{fig:leakage-gaze-vr}
     \end{subfigure}
    \caption{The impact of incomplete privatization of each combination of data streams on user identification rates (yellow). Blue bars display identification rates when all data has undergone privatization, and red bars display identification rates when all of the data is unmodified.}
    \label{fig:privacy-leakage-ir}
\end{figure}

\begin{figure*}
    \begin{subfigure}{0.32\textwidth}
        \includegraphics[width=\textwidth]{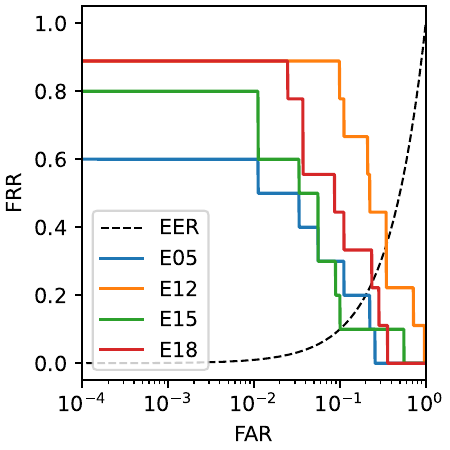}
        \caption{Gaze and head data.}
        \label{leakage-eer-1}
    \end{subfigure}
    \begin{subfigure}{0.32\textwidth}
        \includegraphics[width=\textwidth]{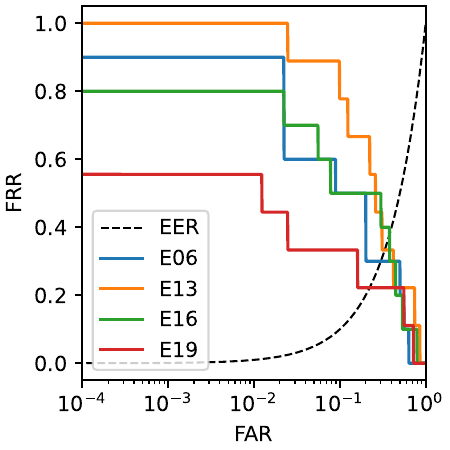}
        \caption{Gaze and hand data.}
        \label{leakage-eer-2}
    \end{subfigure}
    \begin{subfigure}{0.32\textwidth}
        \includegraphics[width=\textwidth]{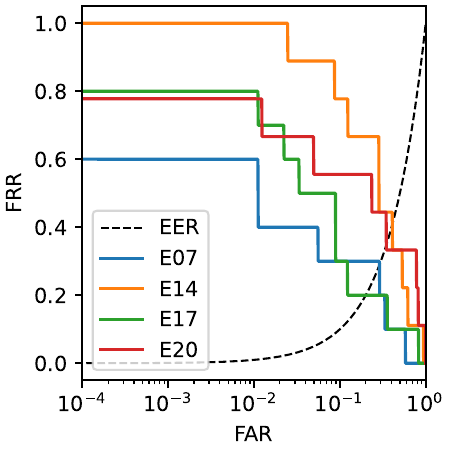}
        \caption{All data streams: gaze, head, and hands.}
        \label{leakage-eer-3}
    \end{subfigure}
    
    \caption{ROC curves showing the effect of patchwork privacy protections to different combinations of data streams on biometric verification.}
    \label{fig:roc-patchwork-privacy}
\end{figure*}

Table~\ref{tab:circumvent} shows the biometric performance of EKYT that was trained on various combinations of privatized and non-privatized data.
Recall that, when head data and hand data are both available, they are either both privatized or both unmodified.
Figure~\ref{fig:privacy-leakage-ir} visually summarizes the difference in identification rates for related test conditions, and Figure~\ref{fig:roc-patchwork-privacy} compares verification performance among the same conditions. 

We first examine biometric performance when both gaze data and head data are available, but privatization is applied only to gaze data (E15) or head data (E18).
Figures~\ref{fig:leakage-gaze-head} and~\ref{leakage-eer-1} visualize the extent to which each unmodified data stream achieves this circumvention.
We previously observed that unmodified gaze data and head data produce similar authentication rates, while privatized gaze data produced better privacy enhancement than privatized head data.
However, when providing privatized gaze data and unmodified head data (E15), EKYT produces biometric authentication rates that are comparable to a condition where no privacy mechanism had been applied at all.
We also observe this effect, albeit to a lesser extent, when EKYT is provided privatized head data and unmodified gaze data (E18).
Despite the demonstrable levels of privacy enhancement afforded to both data streams individually, EKYT was able to leverage the information encoded in the unmodified data stream to successfully authenticate users.

Next, we observe biometric performance when both gaze data and hand data are available, but privatization is applied only to gaze data (E16) or hand data (E19).
Figures~\ref{fig:leakage-gaze-hands} and~\ref{leakage-eer-2} visualize the extent to which each unmodified data stream impacts user authentication rates.
We previously observed that hand data is the least effective data stream for user authentication, and that the privacy mechanism applied to is not particularly effective at obscuring user identities from EKYT---biometric performance did not change significantly between unmodified and privatized hand data.
As a result, attempting to bypass privacy protections applied to gaze data using unmodified hand data (E16) does not significantly improve biometric authentication rates over either the fully privatized data (E13) nor the unmodified baseline performance of hand-only data (E03). 
Because hand data has limited utility as a biometric in this investigation, this attempt to circumvent privacy protections would yield similarly limited levels success.
On the other hand, attempting to circumvent the privacy protections applied to hand data using unmodified gaze data (E19) produces significantly better biometric authentication rates than not only the fully privatized data (E13), but also the test condition where neither data stream was privatized (E06).
While this result is perhaps unsurprising, it highlights that both poorly implemented privacy protections and the complete absence of privacy enhancement can increase the risk of unauthorized user authentication. 

Finally, we observe the biometric performance when both gaze and VR motion data (i.e., head and hand data) are available, but only gaze data (E17) or VR motion data (E20) are privatized. 
Figures~\ref{fig:leakage-gaze-vr} and~\ref{leakage-eer-3} visualize the extent to which each unmodified data stream impacts user authentication rates.
EKYT can effectively circumvent privacy protections when using unmodified VR motion data (E17), producing similar authentication performance as most other experiments that utilize unmodified headset data.
Interestingly, the experiment circumventing privacy protections with unmodified VR motion data (E17) appears to yield significantly better authentication rates than the experiment doing so with unmodified gaze data (E20).
Gaze data is significantly less effective at circumventing privacy protections when VR motion data is privatized (E20)---when EKYT only has unmodified gaze data at its disposal, it produces authentication performance at levels that are comparable to a scenario where all three data streams have been privatized (E14).
In other words, the unmodified gaze data was not effective for user authentication in the presence of privatized VR motion data.
We attribute this finding to the over-representation of VR motion data in EKYT at training time---when training on all three data streams, gaze data only contributes two channels of input and VR motion data collectively contributes nine (see Section~\ref{sec:preprocess} for details).
Because nine of the eleven total channels provided to EKYT were privatized and therefore not informative for biometric authentication in E20, EKYT likely disregarded the user-specific behavior that was only represented in a minority of input channels at training time.
Conversely, this also explains why unmodified VR motion data is highly effective at circumventing privacy protections in E17---the number of input channels that conveyed user-specific behavior outnumbered the number of privatized input channels.

\begin{table}[tb]
  \caption{Biometric performance of EKYT when trained and tested after applying privacy mechanisms to different subsets of data. Performance is averaged across 4-fold cross validation.}
  \label{tab:circumvent}
  \scriptsize%
	\centering%
  \begin{tabular}{%
    l
	*{3}{c}%
    *{2}{r}%
	}
  \toprule
   ID & Gaze & Head &   Hand &   EER (\%, $\downarrow$) &   Rank-1 IR (\%, $\uparrow$) \\ 
  \midrule
    E15 &\private   & \raw &- & 11.9 $\pm$ 3.9 & 70 $\pm$ 14.1 \\ 
    E16 &\private & -  & \raw & 31.4 $\pm$ 10.9 & 45 $\pm$ 26.5 \\ 
    E17 & \private  & \raw  & \raw & 15.8 $\pm$ 5.0 & 72.5 $\pm$ 9.6 \\ 
    E18 & \raw  & \private  & - & 24.5 $\pm$ 3.7 & 57.2 $\pm$ 19.9 \\
    E19 & \raw  & -  & \private & 17.3 $\pm$ 5.4 & 73.1 $\pm$ 25.4 \\
    E20 & \raw  & \private  & \private & 28.1 $\pm$ 6.8 & 48.9 $\pm$ 14 \\
  \bottomrule
    \multicolumn{6}{l}{(\raw)~Unmodified data, (\private)~Privatized data, (-)~Data not used}
  \end{tabular}%
  
\end{table}

\section{Discussion}

The purpose of this investigation is to address the privacy implications of the simultaneous availability of eye tracking and motion data in VR platforms, and how failing to apply privacy protection to all of these data streams may impact the overall risk of user identification.

We first establish that both VR motion data and eye tracking data are suitable for both user verification and identification in our problem setting, and that combining data streams can improve biometric authentication performance when one data stream is more informative than the other.
These findings are consistent with prior work~\cite{Liebers2021} and demonstrate the potentially complementary nature of VR motion data and eye tracking data for user identification.

We then show that existing privacy mechanisms from~\cite{Wilson2024, Nair2023_incognito} can provide demonstrable privacy protections when they are applied to each available data stream.
However, failing to apply these privacy mechanisms to one or more data streams can lead to privacy leakage, which can enable an adversary to circumvent the privacy protections afforded by the protected data streams by leveraging the sensitive information in the unprotected data streams.
In addition, privacy mechanisms that do not impart significant privacy protection (as observed with our hand data) can create a similar vulnerability.
We observed that the level of biometric performance when such privacy leakage is present is comparable to the performance observed when no privacy mechanisms were applied at all, which emphasize the potential ramifications that this vulnerability has on user privacy.
In a real attack scenario, an adversary would not need to know which data streams were unprotected, or if any of the data streams are protected at all.
By emphasizing these implications of our findings, we hope to motivate the development of privacy solutions that are robust to the presence of auxiliary data, especially when that auxiliary data is correlated, synchronized, and complementary in nature.

While this work demonstrates the effect that cross-sensor privacy leakage has on the risk of unauthorized user re-identification in VR platforms, we note that the results of this investigation are heavily influenced by our selection of our privacy mechanisms, authentication model, and data pre-processing strategy, and that some of the trends we observe here may not generalize to other combinations of the three.

First, we utilized two existing privacy mechanisms whose privacy-utility trade-offs and viability for real-time use were validated in prior work.
We observed that the mechanism we applied to VR data was less effective at degrading EKYT's biometric performance overall.
Nair et al.'s~\cite{Nair2023_incognito} privacy mechanism for VR motion data was designed to obfuscate select physical characteristics, which potentially leaves identifying behavioral patterns available to leverage for user identification. 
Although obscuring a user's height and wingspan showed some demonstrable privacy-enhancing effect, EKYT was able to extract other identifiable patterns from the data, potentially from the behavioral patterns that were left unprotected.
This represents an opportunity for advancement in the field of privacy mechanisms for streaming VR data---future work in this direction may explore privacy mechanisms that can obfuscate both physical characteristics and behavioral patterns from VR motion data.

Secondly, we observe that EKYT tends to produce higher biometric performance when gaze data is provided to the model, but also experiences the most degradation in biometric performance when using privatized gaze streams than any other data.
EKYT was originally created and validated for biometric authentication for gaze data, so there may be some artifacts of the architecture design or data processing that may complicate comparisons between its performance on gaze data and non-gaze data.
Although we did not extensively fine-tune our pre-processing of VR motion data to achieve the highest possible biometric performance, we observed relatively high biometric performance on VR motion data.
However, it is possible that biometric performance could additionally be improved by applying sensor fusion techniques to more explicitly define relationships between the data streams.
Pre-processing the input data to explicitly identify user-specific information shared across data streams may help the model learn these correlations more effectively, making the combined data streams more powerful than simply the sum of their individual contributions.

Finally, and most importantly, we note in Section~\ref{sec:some-privacy} that the volume of data that was privatized affected EKYT's ability to effectively learn to extract user-specific behaviors.
Namely, when all eleven input channels are available and all nine channels of VR motion data are privatized, EKYT cannot effectively extract user-specific information from the remaining two channels of unmodified gaze data (E20).
Conversely, when only the two channels of gaze data are privatized, the amount of privacy leakage represented by the nine unmodified channels of VR motion data produce significantly better biometric performance (E17).
While it seems straightforward that privatizing more data generally results in lower levels of privacy leakage, we explicitly note that this is the direct result of the way we provided this input data to EKYT.
If we revised our methodology to enforce that each test condition contains the same number of privatized and unmodified data streams, we anticipate that we would observe less variation in the experiments evaluating the effect of incomplete privacy protection.
This phenomenon highlights one of the many complexities of preserving user privacy in multi-sensor environments like VR.
Although each individual sensor may only contribute a small amount of potentially identifiable information, the sheer volume and variety of data that is collected about a user through these sensors can collectively create an unexpectedly detailed profile of a user over time.
These sensors may perform separate functions to enable different applications in VR, but they can capture potentially complementary information about behavioral idiosyncrasies that may then manifest as privacy vulnerabilities that are difficult to anticipate.

Ideally, a comprehensive privacy solution for VR would involve multiple forms of privacy protection, so that there is no single point of failure that may enable privacy violations.
In addition to applying privacy mechanisms to obfuscate identifiable data and implementing system-level privacy protections mentioned in Section~\ref{sec:prior}, a VR platform can restrict which applications may access these data streams based on their stated purpose.
Applications that do not demonstrate need for sample-to-sample streaming data can be provided abstractions of the data stream which contain less sensitive information that could possibly be exploited.
This strategy has been proposed separately for both gaze data~\cite{davidjohn2021} and gesture-based motion data~\cite{Figueiredo}, and can be combined with other existing techniques to help address the emerging privacy concerns that we highlight in this work.
Ultimately, our work emphasizes the need for privacy solutions that intentionally address the unique privacy considerations of multi-sensor environments, so that privacy protections can be formalized, guaranteed, and experienced consistently.

\section{Limitations and Future Directions}
Although we demonstrate that combining data streams in VR can improve biometric performance when only a portion of data streams are privatized, it is unclear whether it can also allow the model to resolve biometric claims with higher levels of confidence.
Reducing a biometric model's confidence below its decision threshold can be another promising avenue to preventing unauthorized user identification. 
To further investigate whether these data streams contain complementary information for biometrics, future work should report measures of confidence alongside the biometric performance of their models.
Future work in this direction can also improve the explainability of their results by either using statistical, feature-based biometrics models or utilizing explainable AI techniques~\cite{Adadi2018} to provide more insight for deep learning-based biometric models.

Due to the limited availability of data sets that contain both gaze data and VR motion data, this investigation has the limitation of only reporting biometric performance for a relatively small number of subjects.
Compared to our corpus of 38 users, similar studies perform biometric authentication on populations comprising 12~\cite{Liebers2021}, 15~\cite{Olade2020}, 19~\cite{Pfeuffer2019}, and 43~\cite{Wierzbowski2022} people. 
Although our investigation features one of the largest sample sizes available, we acknowledge that having access to a data set with more users can enable a more robust investigation of user identification in VR. 
This work can be expanded by utilizing larger data sets that contain more users who are ideally recorded across multiple sessions, even if these data sets only feature only two of the three data streams that we explored in this work. 
We also limited the scope of this investigation to explore a novel privacy consideration in VR environments because of this limitation, but the insights from this study can reasonably be applied to analogous situations in augmented or mixed reality.

Finally, the results presented in this work may have limited generalizability, as they were produced using a single biometric authentication model working on a single data set, and evaluated only one privacy mechanism per data stream.
While it is clear from our results that the privacy leakage we observe directly impact user authentication rates in this experiment, the extent to which this relationship is observed may vary significantly depending on the authentication model employed, the nature and parameters of the privacy mechanisms that are applied, and any auxiliary pre-processing that a data set may undergo in addition to privacy enhancement. 
Future work in this direction may involve the development and validation of robust, generalizable methods to formally estimate the amount of risk users face as a result of privacy leakage, to enable the development of straightforward ``best practices'' for preserving user privacy applicable to a wider variety of data sets and sensor types.

\section{Conclusion}
We conducted an empirical study exploring a vulnerability in VR systems that stems from the incomplete application of privacy-enhancing techniques to data that is collected on these platforms.
Namely, when privacy protections are applied to only a subset of data streams available on the platform, users can still be vulnerable to identification attacks when unprotected sensor data is used to circumvent them.
We demonstrate that the complementary nature of eye tracking, headset tracking, and hand tracking data allows adversaries to bypass targeted privacy measures and infer user identity with accuracy levels comparable to a scenario where no privacy mechanism is applied at all.
These findings highlight the need for comprehensive privacy-enhancing techniques that account for the collection of synchronized, potentially correlated data streams VR platforms. 
In doing so, future work can advance the state-of-the-art in user privacy for VR ecosystems and contribute to the continued effort to grant users autonomy over their own data in these platforms.

\section{Acknowledgements}
    This material is based upon work supported by the National Science Foundation Graduate Research Fellowship under Grant No. DGE-1840989 awarded to Samantha Aziz. 

\bibliographystyle{plain}
\bibliography{references}

\end{document}